\documentclass{aa}  
\usepackage{graphicx}
\usepackage{txfonts}
\usepackage{tabularx}
\usepackage{appendix}
\usepackage{threeparttable}  
\usepackage{dcolumn}
\usepackage{amsmath}
\usepackage{float}
\usepackage{hyperref}
\usepackage{multirow}
\usepackage[usenames,dvipsnames]{xcolor}
\usepackage{footmisc}
\usepackage{booktabs} 
\usepackage{longtable}
\usepackage{threeparttable}  
\usepackage{comment}

\usepackage[pagewise]{lineno}
\newcommand{\hii}{H{\sc\,ii}}
\newcommand{\uchii}{UC\,H{\sc\,ii}}
\newcommand{\hchii}{HC\,H{\sc\,ii}}

\newcommand{\rms}{r.m.s.}

\newcommand{\ks}{Kolmogorov-Smirnov}

\newcolumntype{d}[1]{D{.}{\cdot}{#1}}

\newcolumntype{.}{D{.}{.}{-1}}
\usepackage[normalem]{ulem}
\usepackage{graphicx}
\usepackage{txfonts}
\begin{document} 
      \title{Dynamic Massive Star Formation:  Radio Flux Variability in \uchii\ Regions}

   \author{A. Y. Yang
          \inst{1,2,5},
          M. A.  Thompson\inst{3},
          J. S.  Urquhart\inst{4}, 
          A. Brunthaler\inst{5}, 
          K. M. Menten\inst{5},
          Y. Gong\inst{6,5},
          Chao-Wei Tsai \inst{1,7,2},
          A. L. Patel \inst{4},
          D. Li \inst{8,1,2},
          W. D. Cotton\inst{9}
         }

   \institute{National Astronomical Observatories, Chinese Academy of Sciences, Beijing 100101, China\\
        \email{yangay@nao.cas.cn}
         \and
    Key Laboratory of Radio Astronomy and Technology, Chinese Academy of Sciences, A20 Datun Road, Chaoyang District, Beijing, 100101, P. R. China 
     \and
    School of Physics and Astronomy, University of Leeds, Leeds LS2 9JT, UK  \\
    \email{M.A.Thompson@leeds.ac.uk}
    \and
    Centre for Astrophysics and Planetary Science, University of Kent, Canterbury, CT2\,7NH, UK 
    \\
    \email{J.S.Urquhart@kent.ac.uk}
    \and Max-Planck-Institut f\"ur Radioastronomie (MPIfR), Auf dem H\"ugel 69, 53121 Bonn, Germany  
    \and Purple Mountain Observatory, and Key Laboratory of Radio Astronomy, Chinese Academy of Sciences, 10 Yuanhua Road, Nanjing 210023, China
\and Institute for Frontiers in Astronomy and Astrophysics, Beijing Normal University, Beijing 102206, China 
\and Department of Astronomy, Tsinghua Univerisity, 30 Shuangqing Road, Beijing 100084, China
\and National Radio Astronomy Observatory, 520 Edgemont Road, Charlottesville, VA 22903, USA \\
             }

  \date{Received September 17, 2024; Revision October 9, 2024; Accepted  October 22, 2024}

 
  \abstract
   {Theoretical models of early accretion during the formation process of massive stars have predicted that \hii\ regions exhibit radio variability on timescales of decades. However, large-scale searches for such temporal variations with sufficient sensitivity have not yet been carried out.   }
   {We aim to identify \hii\ regions with variable radio wavelength fluxes  and to investigate the properties of the identified  objects, especially those with the highest level of variability.}
   {We compared the peak flux densities of 86 ultracompact \hii\ (\uchii) regions measured by the  GLOSTAR and CORNISH surveys and identified variables that show flux variations higher than 30\% over $\sim8$ yr timespan between these surveys. }
   {We found a sample of 38 variable \uchii\ regions,  which is the largest sample identified to date. The overall occurrence of variability  is 44$\pm$5\%, suggesting that variation in \uchii\ regions is significantly more common than prediction.   
   The variable \uchii\ regions are found to be younger than non-variable \uchii\ regions, all  of them meeting the size criterion of hypercompact (HC)\, \hii\ regions.  We studied the 7 \uchii\ regions (the ``Top7'') that show the highest variability with variations > 100\%.  
   The Top7 variable \uchii\ regions are optically thick at 4--8 GHz and compact, suggesting they are in a very early evolutionary stage of \hchii\ or \uchii\ regions. There is a significant correlation between variability and the spectral index of the radio emission. No dependence is observed between the variations and the properties of the sources' natal clumps traced by submillimeter continuum emission from dust, although variable \hii\ regions are found in clumps at an earlier evolutionary stage. }
   {}


 \titlerunning{Flux variability in \uchii\ regions}
\authorrunning{A. Y. Yang, M. A. Thompson, J. S. Urquhart et al}

\maketitle

\section{Introduction}

The canonical picture of the evolution of volumes of gas ionised by massive stars, \hii\ regions, is one of smooth expansion of overpressured ionised gas into a quiescent and largely static ambient interstellar medium \citep[e.g.][]{stroemgren1939, Savedoff1955, Krumholz2009, Churchwell2002ARAA}. Whilst this is likely an accurate description for the late stages of \hii\ region expansion, much younger \hii\ regions must contend with dynamic, high pressure and high density surroundings, coupled with high accretion rate flows that are likely to be gravitationally unstable. Recent models of the early stages of \hii\ region formation have shown that these phenomena can have dramatic time variable effects on the ionization of the \hii\ regions, and consequently their radio brightness \citep[e.g.][]{Krumholz2009, 
 Peters2010a, Peters2010b, Klassen2012, Meyer2017}. \citet{Galvan_Madrid2011MNRAS4161033G} calculated the variability of ultracompact (UC) \hii\ regions from the models presented by \citet{Peters2010a} and showed that the radio brightness of around 10\% of  \uchii\ regions were expected to vary by more than 10\% over a time frame of 10 years. An interesting result of the Galvan-Madrid model is that negative variations are possible (i.e.~the brightness of an UC \hii\ region may \emph{decrease}), although positive variations are statistically more likely.

Observational evidence for radio variability of \hii\ regions has been accumulating over the last two decades, beginning from the measurement of the expansion rate of the archetypal \uchii\ region W3(OH) \citep{Kawamura1998} and of G5.89$-$0.39 \citep{Acord1998} to several observational studies of individual hypercompact (HC) and \uchii\ regions \citep{vanderTak2005AA431993V, Galvan_Madrid2008ApJ674L33G, DePree2015ApJ815123D, DePree2018ApJ863L9D}. Variability in the widths of radio recombination lines has recently been observed in \hchii\ regions within W49A \citep{DePree2020AJ160234D}. Both positive and negative variability have been observed \citep{Galvan_Madrid2008ApJ674L33G, DePree2018ApJ863L9D}. Radio variability seems to be an inherently rare process, with only a handful of confirmed variables --- however only a very small fraction of the $>1200$ known UC or HC \hii\ regions \citep{Urquhart2022MNRAS3389U} have been studied for variability and most of these occur in the most extreme star formation regions in the Galaxy such as Sgr B2 or W49A. 
We currently do not have samples of sufficient statistical power to test the hypotheses of \citet{Galvan_Madrid2011MNRAS4161033G} and fully understand the complex early accretion processes within massive star-forming regions.

In this paper, we present the first statistically significant study of variability in \uchii\ regions based on the data from two recent 5 GHz radio continuum Galactic Plane surveys, the Coordinated Radio and Infrared Survey for High-Mass Star Formation (The CORNISH Survey) \citep{Hoare2012PASP} and the GLOSTAR survey \citep{Brunthaler2021AA651A85B}. Both of these surveys cover the 10\degr $\le l \le$ 60\degr\ region of the Galactic Plane, which contains an estimated $\sim$ 300 \uchii\ regions \citep{Kalcheva2018AA615A103K,Yang2023AA680A92Y}. CORNISH and GLOSTAR data were acquired in similar configurations with the ``classic'' Very Large Array (VLA) and the extended Karl G. Jansky Very Large Array (JVLA)\footnote{The JVLA, originally named ``Extended VLA", available since 2011, is an extensively upgraded version of the ``classic'' VLA, which was commissioned in 1980. Amongst other improvements, the JVLA's much wider bandwidth, more sensitive receivers provide much better sensitivity and instantaneous frequency coverage \citep{Perley2011}.}, respectively, giving them similar angular resolution, a roughly common central frequency, and the same flux calibration scale. 
The CORNISH data were taken between 2006 and 2008 and the GLOSTAR B-array data were taken between 2013 and 2016 (see Sect. \ref{sect:sample}). Thus these data are well  suited to \uchii\ region variability studies as there is an up to \emph{8} year baseline between them. Moreover, the higher sensitivity of GLOSTAR allows a more precise recovery of CORNISH detections. 

 We describe the CORNISH and GLOSTAR surveys and their respective samples of \uchii\ regions in Section \ref{sect:sample}. Our methodology to identify variable \uchii\ regions between the two surveys is detailed in Section \ref{sec:identify}, and the sample of identified variable \uchii\ regions is presented in Section \ref{sect:result}. We discuss the implications of our findings, particularly in the context of the predictions posed by \citet{Galvan_Madrid2011MNRAS4161033G} , in Section \ref{sect:phy_prop} and summarize our conclusions in Section \ref{sect:conclusions}.

\section{The \uchii\ region Sample}
\label{sect:sample}

Both GLOSTAR and CORNISH observations were made in the C-band (4--8\,GHz) in the B and BnA configurations of the ``classic'' VLA (or the upgraded JVLA in the case of GLOSTAR), so the two surveys can be used for a comprehensive comparison of flux densities\footnote{GLOSTAR covered the same areas in the JVLA's B or BnA and D configurations and, to acquire zero spacing information, with the Effelsberg 100 meter radio telescope \cite[for an overview, see][]{Brunthaler2021AA651A85B}. In this study, only the GLOSTAR B/BnA configuration continuum data is considered.} The two flux calibrators 3C286 and 3C48 used both in the CORNISH \citep{Purcell2013ApJS2051P} and the GLOSTAR surveys \citep{Brunthaler2021AA651A85B} are stable with flux density variations below  5\%, as shown in Figure 8 of \citet{Purcell2013ApJS2051P}. 
CORNISH has baseline lengths ranging from approximately $300\,k\lambda$ (i.e., $1\farcs$5) to $2\,k\lambda$ (i.e., $2\arcmin$) in VLA B and BnA arrays and the survey is optimized to detect emission on size scales up to 14\arcsec\ \citep{Purcell2013ApJS2051P}. 
However, GLOSTAR tapered the $uv$ coverage to have baselines larger than $50\,k\lambda$ (i.e., detecting emissions with angular size scales up to $4\,\arcsec$), resulting in a poorer sampling for extended emission compared to the CORNISH survey. 
Due to the differences in spatial filtering, angular resolution, and $uv$ coverage, the comparison of the GLOSTAR and CORNISH radio flux densities is only valid for compact sources defined in the two surveys, as addressed by this study.

GLOSTAR and CORNISH have a considerable overlap across Quadrant I of the Galactic Plane, although the two survey areas are not identical. CORNISH observed a region of the Plane in longitude 10\degr\ $\le \ell \le$ 65\degr \citep[]{Hoare2012PASP} in its B-configuration, while GLOSTAR observed the longitude range  from -2\degr\ $\le \ell \le$ 60\degr plus the Cygnus\,X region, from which the longitude ranges of 2\degr--40\degr\ and 56\degr--60\degr\ are published \citep{Yang2023AA680A92Y,Dzib2023AA670A9D}. Both surveys observed an identical range in latitude ($|b| < 1\degr$), and were imaged at similar central frequencies (5 GHz for CORNISH and 5.8 GHz for GLOSTAR), with similar angular resolutions of $1\farcs5$ for CORNISH and $1\farcs0$ for GLOSTAR. The observed noised levels of  GLOSTAR survey are lower, with a median r.m.s. of $\rm \sim 0.08\,mJy\,beam^{-1}$, compared to  $\rm \sim 0.4\,mJy\,beam^{-1}$ for CORNISH data. 
In terms of the temporal baseline between the two surveys, CORNISH was observed between 2006 July 12 and 2008 February 4. The GLOSTAR data used in this study were taken between 2013 September 29 and 2016 October 28. The mean separation between the two surveys is approximately 8 years.

Within the overlap region of the published GLOSTAR B-configuration data and the CORNISH coverage,  219 \hii\ regions were detected in the high-reliability catalog (SNR>7$\sigma$), including 184 \uchii\ regions \citep{Purcell2013ApJS2051P,Kalcheva2018AA615A103K}.  
Among the CORNISH \uchii\ regions, 56 \uchii\ regions are compact (defined as sources with angular sizes $\leq$1.8\arcsec\ in \citealt{Purcell2013ApJS2051P}) and not associated with over-resolved emission structures. 
In the overlap region, GLOSTAR detected 353 \hii\ regions due to the better sensitivity, and 227 of them are compact (i.e., \uchii\ regions), defined as sources with $Y_{\rm factor}<2$ in GLOSTAR \citep{Dzib2023AA670A9D,Yang2023AA680A92Y}, where the $Y_{\rm factor}$ is defined as the numerical ratio between integrated flux density (or flux) and peak flux density (or brightness). 
The identification process of variable \hii\ regions is based on comparing peak flux densities in the two \uchii\ region samples. 
The peak flux densities are used to compare for the following reasons. First, the integrated flux densities of \uchii\ regions are affected by different configurations and $uv$ coverage of VLA. This is because the \uchii\ regions are not perfectly unresolved and have a hierarchical structure of ionized gas in complex environments and their observed angular sizes are thus always similar to beam sizes under different beams, as seen in \citealt{Yang2019ApJS24118Y,Yang2021AA645A110Y}. The flux density of compact sources with sizes comparable to the beam size would be affected by the $uv$ coverage, as seen in Figure 15 of \citet{Plunkett2023PASP135c4501P}. Additionally, the peak flux averaged over the beam rather than the entire source is commonly used for the measurements of \uchii\ regions in complex regions \citep[e.g.,][]{Wood1989ApJS69831W}. 
Also, many \uchii\ regions are detected by GLOSTAR above 7$\sigma$ but not detected by CORNISH above 7$\sigma$, which makes it difficult to accurately measure their integrated flux densities from CORNISH, as outlined in Sect.\,\ref{sec:identify}. 

\section{Identification of Variable \hii\ Regions} 
\label{sec:identify}

\begin{figure*}[!htp]
 \centering
\includegraphics[width = 0.95\textwidth]
{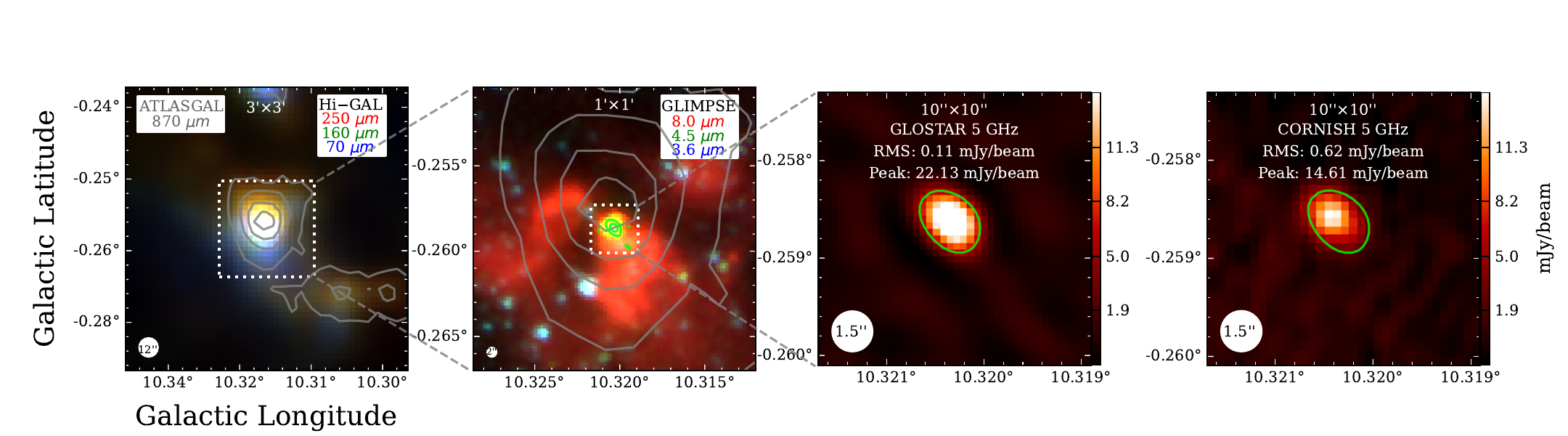} \\
 \vspace{-1.0mm}
(a) Variable \hii\ region G010.3203-00.2578 in Criterion (1) \\
\includegraphics[width = 0.95\textwidth]
{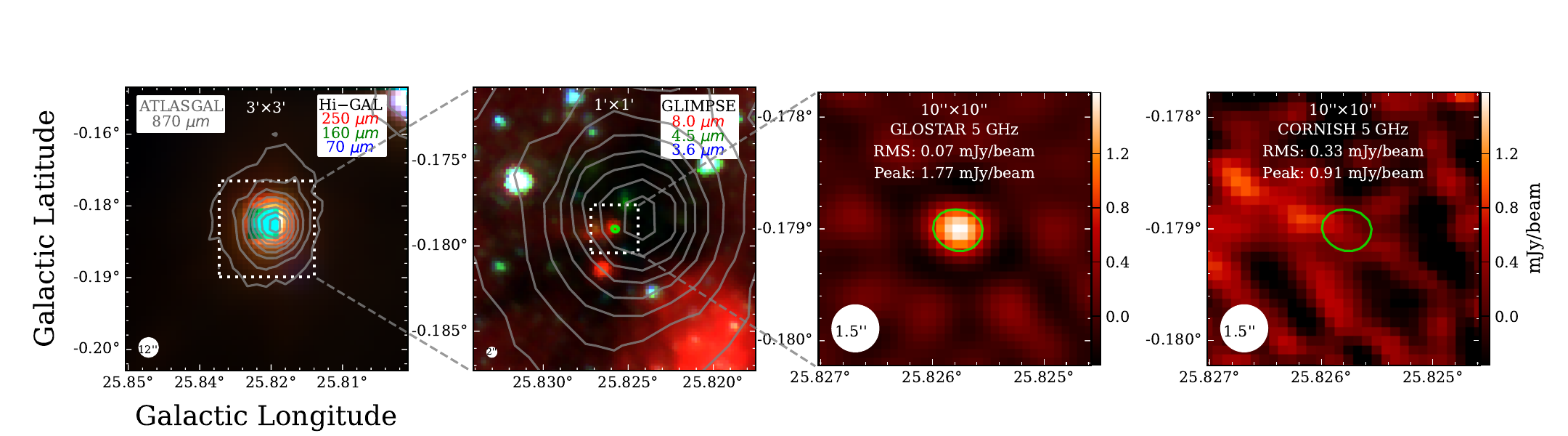} \\
 \vspace{-1.0mm}
(b) Variable \hii\ region G025.8258-00.1790 in  Criterion (3)\\
\includegraphics[width = 0.95\textwidth]
{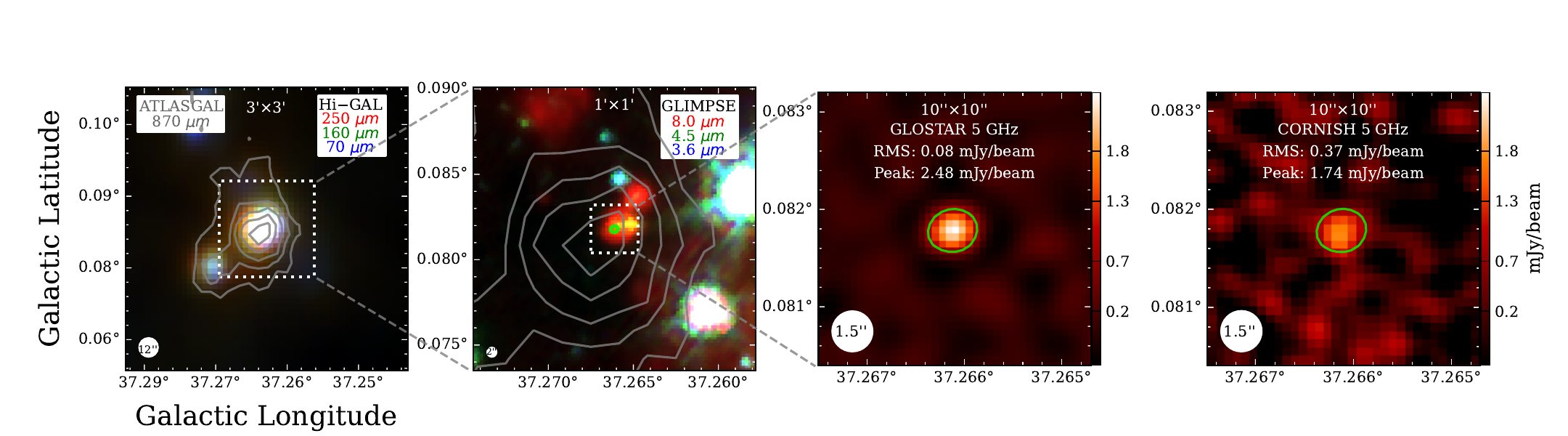} \\
 \vspace{-1.0mm}
(c) Variable \hii\ region G037.2661+00.0818 in Criterion (3) \\
\includegraphics[width = 0.95\textwidth]
{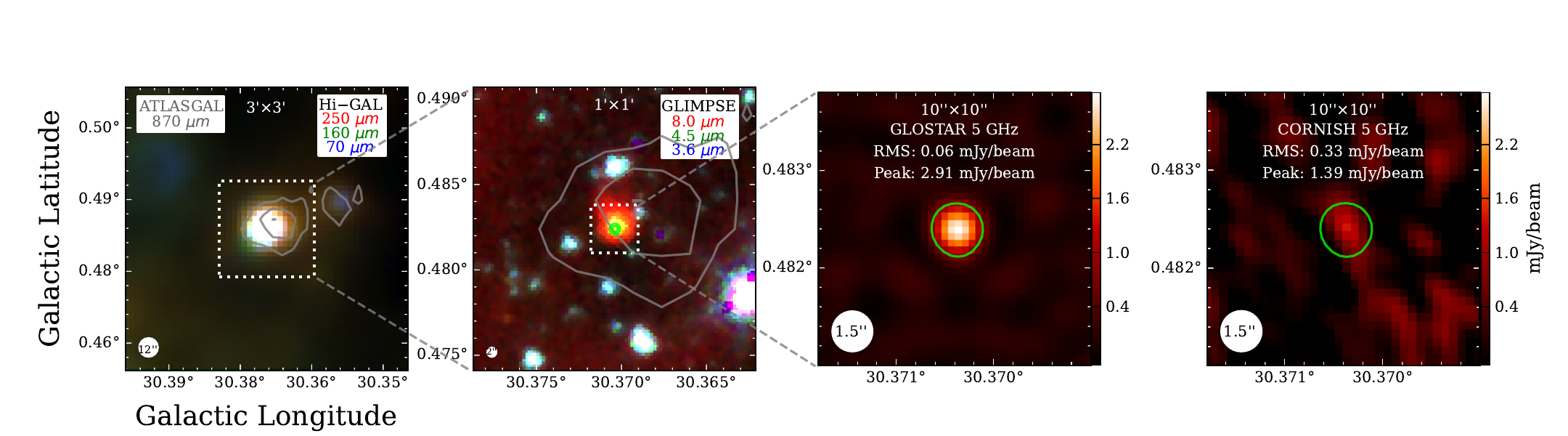} \\
 \vspace{-1.0mm}
(d) Variable \hii\ region G030.3704+00.4824 in Criterion (3) \\

\caption{Examples of multi-band images of variable \hii\ regions identified by Criteria (1)--(3) in Sect.\,\ref{sec:identify}. We noted that no source meets the Criterion (2). 
From top to bottom: (a) An example of a variable \hii\ region that meets Criterion (1); (b)-(d) Example of variable \hii\ regions that meets Criterion (3): (b) using 3$\sigma$ as the upper limit of the CORNISH peaks to estimate the variation if 3$\sigma$ is higher than the local peak; (c) and (d)  using the local peak of the source to estimate the variation if the local peak is higher than 3$\sigma$; 
From left to right: three-colors composition image of Hi-GAL (red = 250\,$\mu m$, green = 160\,$\mu m$, blue = 70\,$\mu m$), GLIMPSE (red = 8.0\,$\mu m$, green = 4.5\,$\mu m$, blue = 3.6\,$\mu m$), radio image of GLOSTAR and CORNISH at 5\,GHz. 
The noise level of CORNISH is $\rm 0.33\,mJy\,beam^{-1}$. 
 The white circles in the lower-left corner of each image show the FWHM beams of Hi$-$GAL\,(12\arcsec\ at 160\,$\mu m$), GLIMPSE\,(2\arcsec), GLOSTAR\,(its B-configuration smoothed to 1.5\arcsec) and CORNISH\,(1.5\arcsec). The gray contours in the left two panels show the ATLASGAL emissions at 870\,$\mu m$. The lime contours in the middle and right panels show the radio emission of GLOSTAR. The image size is indicated on the top of each figure centered at the coordinates of each \hii\ region. 
 The GLOSTAR images have been smoothed to match the beam size of 1.5\arcsec\ in CORNISH.  }  
 \label{fig:example_var_hii}

 \end{figure*}


We defined a sample of variable \uchii\ regions based on the ratios of peak flux densities measured in the GLOSTAR and CORNISH surveys,  that are not over-resolved in complex emission regions. 
To allow a meaningful comparison of the peak flux densities of compact sources, the GLOSTAR images were  smoothed to match the CORNISH beam size of $1\farcs5$. The slight differences in observational parameters between the two surveys result in differences in the Fourier-space sampling \citep{Perley1989ASPC6P}, and the center frequencies. We conservatively estimate that any flux changes under the  10\% level might be questionable, as discussed by \citet{Galvan_Madrid2011MNRAS4161033G}. 
To ensure that we only consider true variability in the sample, we define an \uchii\ region as a variable, if its peak flux densities measured in the data from the two surveys have a greater than 30\%\ variation fraction (defined below).

A \uchii\ region is considered to be variable if it satisfies any of the following criteria:

\begin{enumerate}
\item[] (1) The peak flux densities between GLOSTAR ($S_{\rm peakG}$) and CORNISH ($S_{\rm peakC}$) exhibit a variation fraction  $|f_{\rm var}|$ greater than 30\%, where $ f_{\rm var} = (S_{\rm peakG}-S_{\rm peakC})/S_{\rm peakC}$,  or the flux ratio $S_{\rm peakG}/S_{\rm peakC}$ greater than 1.3 or less than 0.7. \\
\item[] (2) A source is detected in CORNISH above 7$\sigma$ but is not detected by GLOSTAR at the 7$\sigma$ level; $\sigma$ is the local \rms\ noise level, with $|f_{\rm var}|>30\%$.\\


\item[] (3) A source is detected in GLOSTAR above 7$\sigma$, but is not detected by CORNISH at 7$\sigma$, and the GLOSTAR peak $S_{\rm peakG}$ is higher than the local CORNISH detection limit (i.e., $3\,\sigma$), with $|f_{\rm var}|>30\%$. \\

\end{enumerate}
Among the 227 \uchii\ regions in GLOSTAR (see Section\,\ref{sect:sample}), 56 of them are detected by CORNISH above 7$\sigma$.  Criterion (1) results in 22 (39.2\%, 22/56) variable  \uchii\ regions detected by GLOSTAR and CORNISH above 7$\sigma$, as shown in the top-panel of Fig.\,\ref{fig:example_var_hii}. As all compact CORNISH \uchii\ regions are detected in the GLOSTAR B-configuration images \citep{Dzib2023AA670A9D,Yang2023AA680A92Y}, no source meets Criterion (2). 
From 171 \uchii\ regions detected by GLOSTAR above 7$\sigma$ but not detected by CORNISH above 7$\sigma$, we cannot determine the flux variations for 141 sources  (82\%,141/171) using the two surveys as these sources are below the 3$\sigma$ local detection level of CORNISH ($\sigma$ is the local r.m.s. of CORNISH) and their flux densities in CORNISH cannot be estimated. 
This leaves a sample of 30 \uchii\ regions higher than 3$\sigma$ local detection level of CORNISH that can be used to measure $f_{\rm var}$ in Criterion (3). 
To calculate the  variations of compact \hii\ regions meeting Criterion (3), we measured their peak flux densities and local \rms\ noise level ($\sigma$) from the CORNISH images at the position of each source. 
If the local peak flux density value is less than or equal to 3$\sigma$, the 3$\sigma$ is used as the upper limit of the CORNISH peak, as illustrated in panel (b) of Fig.\,\ref{fig:example_var_hii}. 
If the local peak value is greater than 3$\sigma$, the local peak value is considered the CORNISH peak, as illustrated in panel (c) and panel (d) of Fig.\,\ref{fig:example_var_hii}. 
The determined CORNISH peak was used to estimate the flux density variation $f_{\rm var}$.
This results in 16 out of the 30 in Criterion (3), which have variation fractions $|f_{\rm var}|>$\,30\%.

\section{Results}
\label{sect:result}

\begin{figure}[!htp]
 \centering
\includegraphics[width = 0.45\textwidth]
{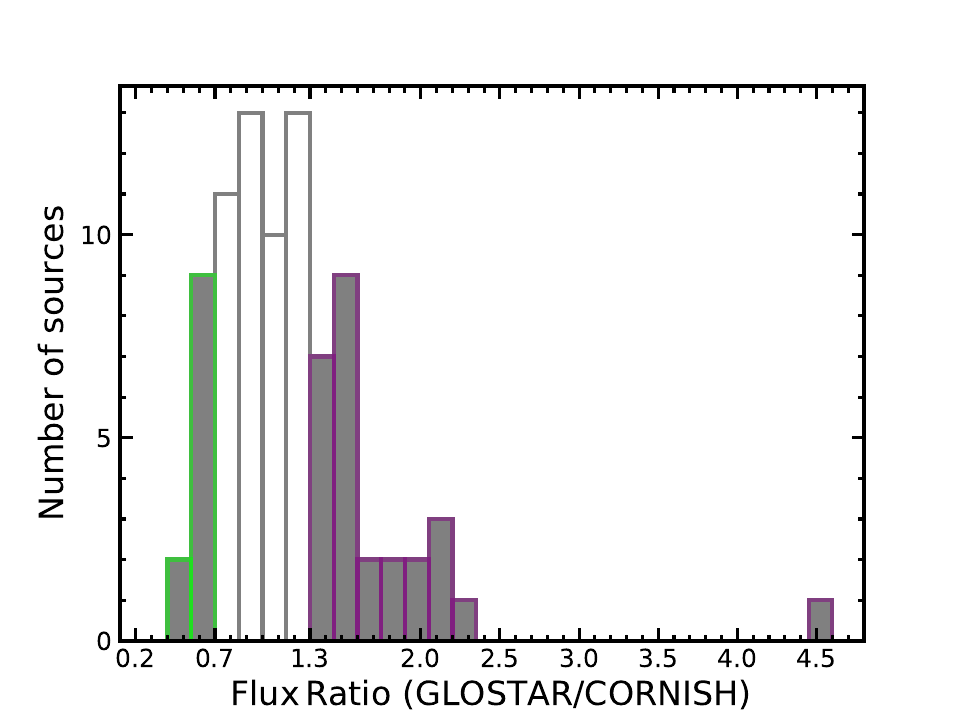} \\
\caption{Distribution of flux ratio between GLOSTAR and CORNISH for the 86 \uchii\ regions (white shaded) that can be assessed using Criteria (1)-(3). This gives 38 variable \uchii\ regions (grey shaded) with flux variation fraction $|f_{\rm var}|>$30\%, which includes purple area for those with the flux ratio > 1.3 (or $f_{\rm var}>30\%$) and the lime area for those with flux ratio < 0.7 (or $f_{\rm var}<-30\%$). The bin size is 0.15. }  
 \label{fig:distr_flux_ratio_all}
 \end{figure}

From the criteria outlined in the previous section, we obtain a sample of 38 \uchii\ regions with variation fraction $|f_{\rm var}|>30$\%, including 22 sources from Criteria (1) and 16 sources from Criteria (3). 
Fig.\,\ref{fig:distr_flux_ratio_all} shows the distribution of flux ratios between GLOSTAR and CORNISH for all 86 sources (white) that can be assessed using the three criteria and the 38 variable \uchii\ regions (grey shaded) identified by this study. We display the GLOSTAR and CORNISH images at 5 GHz for the 38 variable sources in Fig.\,\ref{fig:38_variables_radio_images}. 
Table\,\ref{tab:7_most_var_HII} lists the physical properties of the 7 most variable \hii\ regions, and the full table for the 38 variable sources is available online. The source angular size $\theta_{\rm S}$ given in Column 8 is derived from the geometric mean of the major and minor axes $\sqrt{\theta_{major}\times\theta_{minor}}$. The GLOSTAR peaks $\rm S_{peakG}$, the major ($\theta_{major}$) and minor ($\theta_{minor}$) axis are measured from the GLOSTAR smoothed images by 2d Gaussian fit using AEGEAN \citep{Hancock2012MNRAS4221812H}, and the deconvolved sizes $\theta_{dcon}=\sqrt{\theta_{S}^{2}-beam^{2}}$ are determined by the strategy used in previous studies for compact and young \hii\ regions \citep[e.g.,][]{Purcell2013ApJS2051P,Yang2021AA645A110Y,Patel2023MNRAS5244384P,Patel2024MNRAS5332005P}. 
The distances of natal clumps in Table\,\ref{tab:7_most_var_HII} , together with their bolometric temperature, luminosity, and mass, are measured by the $870 \mu m$ APEX Telescope Large Area Survey of the Galaxy \citep[ATLASGAL][]{Schuller2009AA} by \citealt{Urquhart2018MNRAS4731059U,Urquhart2022MNRAS3389U} and Hi-GAL  \citealt{Elia2021MNRAS5042742E}. 
The remaining properties in Table\,\ref{tab:7_most_var_HII} are obtained from GLOSTAR \citep[e.g.,][]{Brunthaler2021AA651A85B,Dzib2023AA670A9D,Yang2023AA680A92Y} and CORNISH \citep{Hoare2012PASP,Purcell2013ApJS2051P,Kalcheva2018AA615A103K}.

\subsection{The variable sample}
\label{sect:the_var_sample}
Among the 38 variables, 22 sources show a variation fraction >50\%. Nine sources exhibit the highest variability, with a variation fraction of $\geq$ 100\%, indicating a peak flux density ratio between GLOSTAR and CORNISH greater than 2.0 or less than 0.5. 

\citet{Kalcheva2018AA615A103K} compared images from CORNISH and 6 cm data compiled by \citet{White2005AJ} from various VLA survey efforts, and found a total of 9 variable \uchii\ regions. Three of these sources, namely G011.9786-00.0973, G014.5987+00.0198, and G025.7156+00.0487, were also identified as variable sources by us. The flux density ratios they find for these three sources are roughly twice as large as the values we find. This is expected since the time baseline in \citet{Kalcheva2018AA615A103K} is 15 years compared to 8 years in this work. Three other sources, G16.3913-00.1382, G30.7579+00.2042, and G30.7661-00.0348, were discarded by us because they are extended. For the remaining three sources, namely G11.0328+00.0274, G23.4553-00.2010, and G37.7347-00.1128, we find no significant changes between the CORNISH and GLOSTAR observations, even though the latter two are reported with very large changes in \citet{Kalcheva2018AA615A103K}. However, these three sources were not detected in the earlier 6 cm data from \citet{White2005AJ}, and the variability estimates were based on upper limits. Since the CORNISH and GLOSTAR observations are much more similar than the earlier 6 cm observations with a lower spatial resolution, we do not consider these three sources as variable. Furthermore, we noted that our sample of variables comprises 8 out of the 13 variable \hii\ regions reported by \citet{Dzib2023AA670A9D} and \citet{Yang2023AA680A92Y} with 5 variable sources excluded because they are over-resolved in complex emission structures, as outlined in Section\,\ref{sect:sample}. 

\begin{figure}[!htp]
 \centering
\includegraphics[width = 0.45\textwidth]
{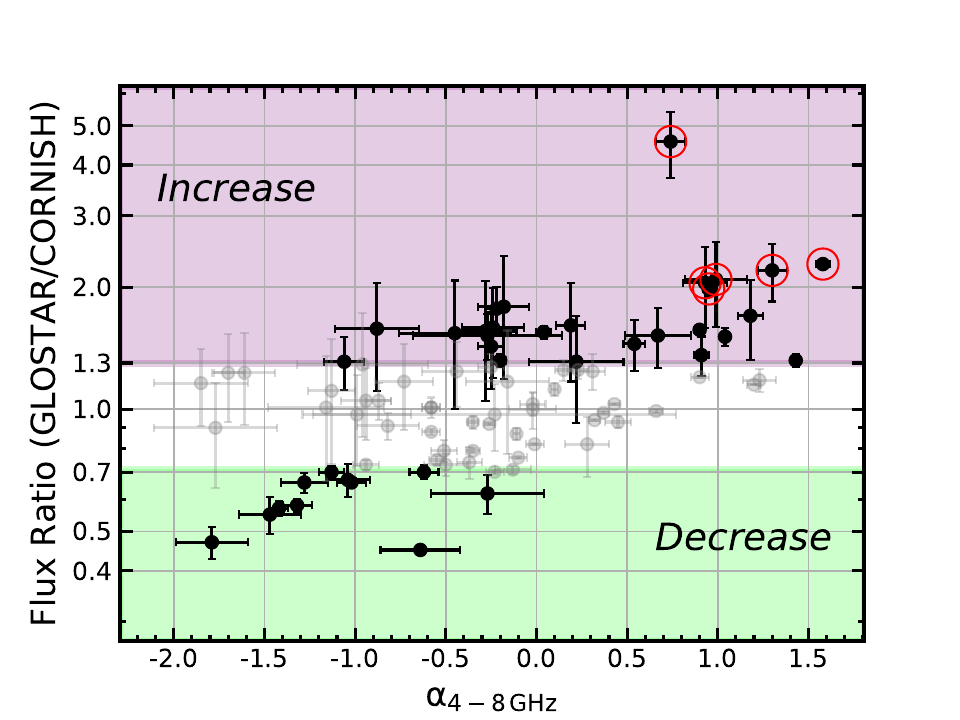} \\ 
\caption{The flux ratio between GLOSTAR and CORNISH as a function of the GLOSTAR spectral index at 4--8 GHz of the 86 sources, including 38 variables (black dots) with $|f_{var}|>30\%$ and the remaining 48 sources (grey dots) with $|f_{var}|<30\%$, as defined in Sect.\,\ref{sec:identify}. The purple shaded area includes the variables with flux increase and the lime shaded area includes the variables with flux decrease, as outlined in Fig.\,\ref{fig:distr_flux_ratio_all}. The red circles display the Top7 variable \uchii\ region (see Sect.\,\ref{sect:the_var_sample} for details) with 6 having reliable spectral indices listed in Table\,\ref{tab:7_most_var_HII}.} 
 \label{fig:alpha_vs_fluxratio_hii}
 \end{figure}

The spectral indices between 4\,GHz and 8\,GHz for the 38 variables were measured by the GLOSTAR survey \citep{Yang2023AA680A92Y}. From Fig.\,\ref{fig:alpha_vs_fluxratio_hii}, we can see that the 38 variables show a wide range of spectral indices from $-$1.8 to +1.6. It is evident, that all sources with a spectral index above $\alpha_{4-8\,GHz}\gtrsim-0.1$, i.e. consistent with optically thin or thick free-free emission, are rising in flux density in the eight years between the CORNISH and GLOSTAR observations. On the other hand, the flux densities of sources with spectral indices below -0.1 can be increasing or decreasing.  Because of their spectral index, these sources are likely to contain significant non-thermal emission (e.g. from  a variable radio jet) as discussed in \citet{Yang2023AA680A92Y}. Considering the larger uncertainty in the spectral index in GLOSTAR for weak sources \citep{Yang2023AA680A92Y}, additional multi-band observations are required to accurately determine their spectra. However, such analysis is beyond the scope of the current study.  

Among the most variable sources with flux density ratios below 0.5 or above 2, two sources showed a decrease in flux density between the CORNISH and  GLOSTAR observations. However, these two sources (G023.1974-00.0005 and G028.4518+00.0029) have also a negative spectral index $\alpha_{4-8\,{\rm GHz}}<-0.5$ as shown in Fig.\,\ref{fig:alpha_vs_fluxratio_hii} and thus are likely to be contaminated by radio jets in the star formation regions, as discussed in \citet{Yang2023AA680A92Y}. The non-thermal radio jets with negative spectral index have been reported in \citet{Anglada2018AARv263A}, and \citet{Obonyo2024MNRAS5333862O} found 50\% of radio jets from massive protostars emit non-thermal radiation based on the data from the SARAO MeerKAT Galactic Plane Survey \citep{Goedhart2024MNRAS531649G}. 
The remaining 7 variables exhibiting the highest variability (hereafter Top7) show positive spectra and have a flux density increase from CORNISH to GLOSTAR, as shown in Fig.\,\ref{fig:alpha_vs_fluxratio_hii}.




%

Out of the 227 \uchii\ regions detected in the GLOSTAR survey, 86 sources are suitable for analyzing variability based on Criterion (1)-(3), resulting in 38 variable sources. 
This gives an overall occurrence of variability in our \uchii\ region sample of 44$\pm$5\% (38/86). The uncertainties of the fractions are calculated from the standard error of binomial distribution of fraction $f$,
that is, $\sqrt{f*(1-f/(num))}$ where $num$ is the sample size of 86 in this study. 
Taking into account the uncertainty, the fraction of \uchii\ regions with radio continuum flux variations appears to be somewhat consistent with the fraction of infrared continuum variability of 32.5\% observed in the sample of 2230 star-forming clumps studied by \citet{Lu2024ApJS27244L} based on the NEOWISE database \citep{Mainzer2011ApJ743156M}.
The observed fraction of 44\% is larger than the predicted of $\sim$10\% of observed \hii\ regions should have detectable flux variations on time-scales of $\sim$10 yr in \citet{Galvan_Madrid2011MNRAS4161033G}.
In the simulation of radio continuum flux variation of \uchii\ regions in \citet{Galvan_Madrid2011MNRAS4161033G} for Run A with a single sink, they  predicted: 16.7$\pm$2.9\%, 6.8$\pm$1.9\%, and 4.7$\pm$1.6\% are expected to have flux increments larger than 10\%, 50\%, and 90\%, respectively, for two observations separated by 10\,yr. The more realistic Run B for a single sink with additional collapse events and
clusters formation in \citet{Galvan_Madrid2011MNRAS4161033G} predicts a smaller fraction of variable \uchii\ regions compared to Run A: 6.9$\pm$1.6\%, 0.3$\pm$0.3\%, and 0.0\% of them are expected to have flux increments larger than 10\%, 50\%, and 90\%, respectively. 
In this study, we found much higher fractions of 50.0$\pm$5\%, 23.3$\pm$4\%, and 8.1$\pm$3\%,  showing flux increments exceeding 10\%, 50\%, and 90\%. Therefore, variable \uchii\ regions could be significantly more common than predicted. 
Given that the models used in \citet{Galvan_Madrid2011MNRAS4161033G} to simulate the radio flux variation of \uchii\ regions do not include the effect of radio jets, the higher fraction of variable \uchii\ regions found in this work support that the variable sample is contaminated by radio jets in star formation regions, as discussed above. 
It is important to note that the difference in wavelength between simulations (2\,cm) in  \citet{Galvan_Madrid2011MNRAS4161033G}  and observations (6\,cm) in this study may contribute to the inconsistency in variation fractions. 
This is because the origin of observed variations could be due to the density structures of young \hii\ regions and their intermediate to large optical depths, as mentioned in \citet{Peters2010a,Peters2010b} and \citet{Galvan_Madrid2011MNRAS4161033G}. The young \hii\ regions are more optically thick with higher optical depths at the observation at 6\,cm compared to the simulation at 2\,cm. 
A comparison between simulations and observations in the same frequency is thus needed to understand the discrepancy. 
However, such analysis is beyond the scope of this study. 

\setlength{\tabcolsep}{2.0pt}
\begin{table*}
\centering
\caption{ \it \rm The 38 variables
 observed in the GLOSTAR and CORNISH for $10\degr<\ell<40\degr$ and  $56\degr<\ell<60\degr $ with $|b| < 1\degr$.  Columns 1-5 correspond to the  Galactic name, Galactic longitude $\ell$ and latitude $b$, the right ascension (RA), and declination (Dec) of the GLOSTAR sources. Columns 6-7 list the peak flux densities and the errors from GLOSTAR ($\rm S_{peakG}\pm dS$) and CORNISH ($\rm S_{peakC}\pm dS$) under the same beam size of 1.5\arcsec. 
 Sources with $\theta_{S}> 1.5\arcsec$ (i.e., the beam size) in Column 8 are considered resolved, and their deconvolved sizes $\theta_{dcon}$ are determined in Column 9. Sources with $\theta_{S}\leq 1.5\arcsec$ (i.e., the beam size) are considered unresolved, and their $\theta_{dcon}$ and $diam$ are considered the upper-limit values, as marked by $\star$ in Column 1, and the upper-limit of $\theta_{dcon}$ is set to be 0\farcs 32, which corresponds to one standard deviation of the distribution of beam sizes from the CORNISH survey \citep{Purcell2013ApJS2051P}. Column 12 gives the variation fraction, as outlined in Sect.\,\ref{sec:identify}.
The spectral indices $\alpha$ and their errors $d\alpha$ are listed in Column 13.  }
 \begin{tabular}{lcccccccccccccc}
\hline
\hline
GLOSTAR B-conf & $\ell$ & $b$ & $RA$ & $Dec$ & $S_{\rm peakG}\pm dS$ & $ S_{\rm peakC}\pm dS$  & $\theta_{S}$  & $\theta_{dcon}$ & $ diam$ & $ Dist.$ & $f_{\rm var}$ & $\alpha\pm d\alpha$\\
Gname & $\degr$ & $\degr$ &$\degr$ & $\degr$ & $\rm mJy/beam$    & $\rm mJy/beam$    & $\arcsec$  &  $\arcsec$  & pc & kpc & \% & \\
\hline
(1) & (2) & (3) & (4) & (5) & (6) & (7) & (8) & (9) & (10) & (11) & (12) & (13)   \\
\hline
G010.4629+00.0299$\star$ & 10.4629 & 0.02988 & 272.15194 & -19.87086 & 11.15$\pm$0.23 & 2.44$\pm$0.45 & 1.4 & 0.32  & 0.013 & 8.5 & 357.0 & 0.74$\pm$0.08 \\
G010.4724+00.0274 & 10.47236 & 0.02742 & 272.1591 & -19.86378 & 50.96$\pm$0.22 & 22.34$\pm$2.0 & 1.67 & 0.73 & 0.030 & 8.5 & 128.1 & 1.58$\pm$0.04 \\
G010.8851+00.1225$\star$ & 10.8851 & 0.12249 & 272.28323 & -19.45668 & 3.51$\pm$0.11 & 1.71$\pm$0.38 & 1.32 & 0.32 & 0.004 & 2.8 & 105.2 & 0.93$\pm$0.12 \\
G011.9039-00.1412 & 11.90387 & -0.14122 & 273.04769 & -18.69153 & 21.97$\pm$0.13 & 11.11$\pm$1.1 & 1.51 & 0.21 & 0.004 & 3.7 & 97.7 & 0.95$\pm$0.03 \\
G014.6095+00.0125$\star$ & 14.60949 & 0.01246 & 274.26125 & -16.24118 & 3.63$\pm$0.08 & 1.76$\pm$0.39 & 1.49 & 0.32  & 0.017 & 11.2 & 106.0 & $-$ \\
G025.7094+00.0438$\star$ & 25.7094 & 0.04377 & 279.51311 & -6.40431 & 4.93$\pm$0.09 & 2.24$\pm$0.36 & 1.4 & 0.32  & 0.016 & 10.2 & 120.1 & 1.30$\pm$0.08 \\
G030.3704+00.4824$\star$ & 30.37037 & 0.48241 & 281.26131 & -2.05938 & 2.91$\pm$0.06 & 1.39$\pm$0.33 & 1.39 & 0.32 & 0.014 & 9.2 & 109.3 & 0.99$\pm$0.17 \\
\hline
\hline
\end{tabular}
\begin{tablenotes}
\item{ The symbol $``-"$ in Column (13)  means that the $\alpha$ is unreliable. Only the Top7 variables are provided here. The full table of the 38 variables is available in electronic form at CDS.}  
\end{tablenotes}
\label{tab:7_most_var_HII}
\end{table*}

\section{ Properties of variable \uchii\ regions}
\label{sect:phy_prop}
\subsection{\uchii\ region properties}
\label{sect:hii_prop}
Fig.\,\ref{fig:alpha_vs_fluxratio_hii}, shows a strong correlation between the spectral index $\alpha_{4-8\,GHz}$ and the flux ratio between GLOSTAR and CORNISH, with a Spearman's rank coefficient $\rho$ = 0.6 and $p$-value $\ll$ 0.001, indicating that variable sources with steeper spectral indices (both negative and positive) have larger variations. 

\begin{figure}[!htp]
 \centering
\includegraphics[width = 0.45\textwidth]
{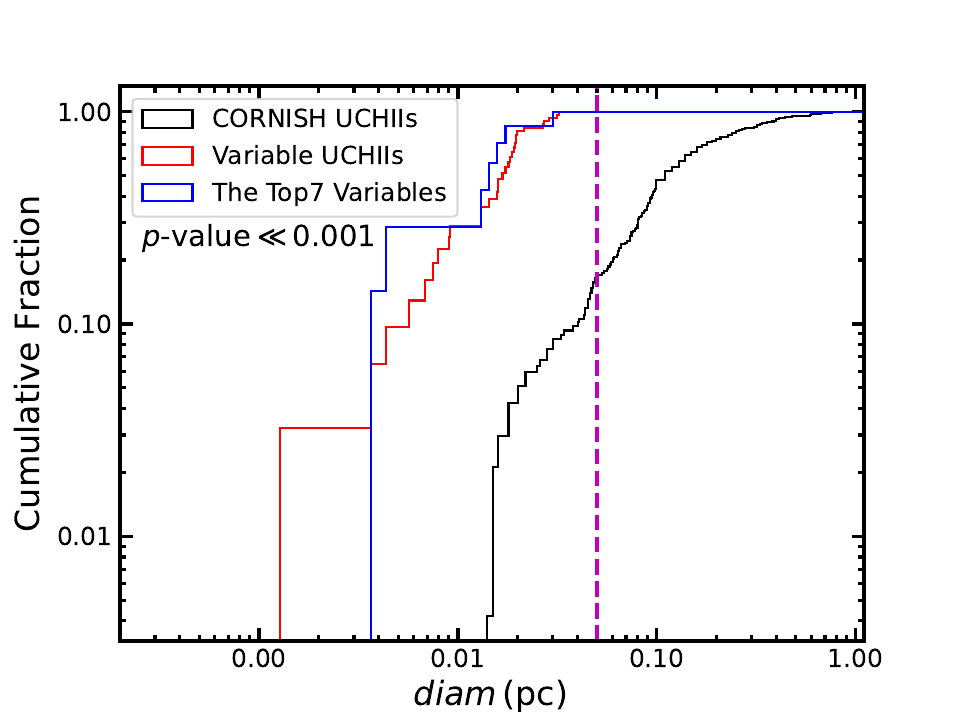} \\
\includegraphics[width = 0.45\textwidth]{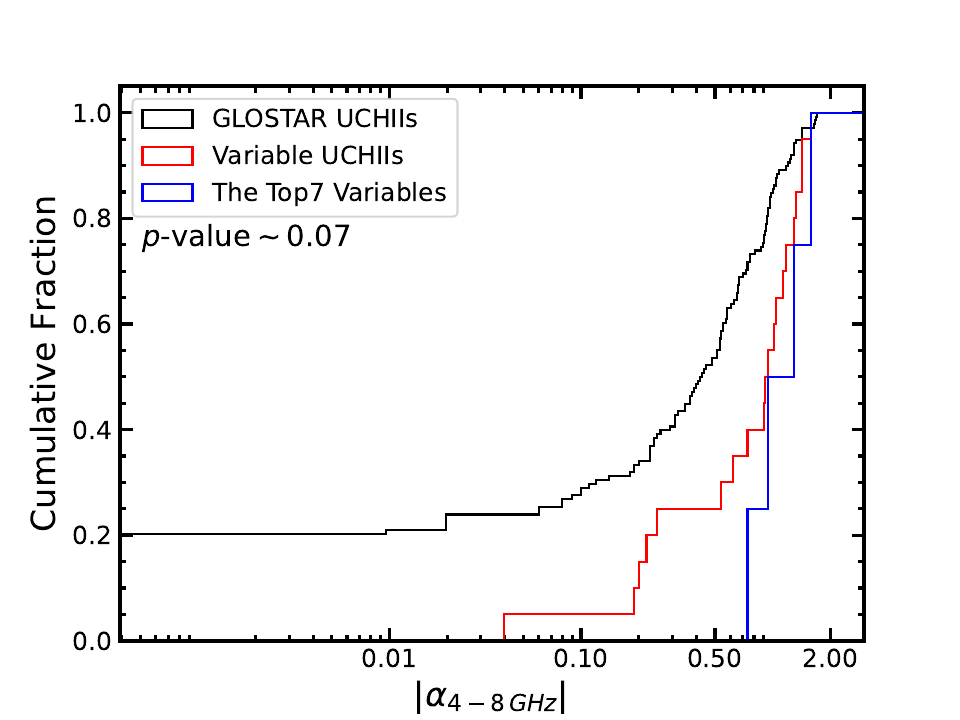} \\

\caption{Cumulative distributions of the physical diameter (top-panel) and the spectral index $\alpha_{4-8\,GHz}$ (bottom panel) for the Top7 (blue) and variable \uchii\ regions (red) of this work, as well as for the \uchii\ regions (black) from the GLOSTAR \citep{Yang2023AA680A92Y} or CORNISH \citep{Kalcheva2018AA615A103K}.
32 out of the 38 variables \uchii\ regions have their distances measured in \citet{Urquhart2018MNRAS4731059U}. 
The magenta line indicates diam=0.05\,pc, i.e., the size criterion of the \hchii\ regions. This indicates that 100\% of variable sources in this study meet the size criterion of \hchii\ regions. 
As the uncertainty in spectral index $\sigma_{\alpha}$ is large for weak sources in GLOSTAR \citep{Yang2023AA680A92Y}, we limit the sources with $\sigma_{\alpha}<0.1$ here for $\alpha$ comparison. }
 \label{fig:diam_alpha_var_uchii}
 \end{figure}

This trend, where steeper spectral indices are associated with higher variations, is also significant when comparing against the entire sample of GLOSTAR \uchii\ regions with measured in-band spectral indices. In the lower panel of Fig.\,\ref{fig:diam_alpha_var_uchii}, we present cumulative distributions of the spectral indices for all GLOSTAR \uchii\ regions (GLOSTAR is used here due to it has the measured spectral index), our variable sample, and the Top7 variable objects. A K-S test allows us to reject the null hypothesis that variable \uchii\ regions and GLOSTAR \uchii\ s are drawn from the same parent population ($p$-value $=$ 0.07).

We also observe significant evidence for higher variability in smaller \uchii\ regions, in the top panel in Fig.\,\ref{fig:diam_alpha_var_uchii} where we plot the cumulative distributions of physical diameters $diam$ for the whole CORNISH \uchii\ sample (CORNISH is used here due to it has the measured physical diameters for \uchii\ regions), the variable \uchii\ regions and our Top7 most variable sources in this work. 
A \ks\,(K-S) test again allows us to reject the null hypothesis that the variable \uchii\ sample and the CORNISH \uchii\ sample are drawn from the same parent population with a $p$-value $\ll$ 0.001. 
We note that many variable \uchii\ regions are unresolved and have only upper limits to their diameters --- however, these upper limits show that over 50 \% of them are consistent with the expected diameters of \hchii\ regions (i.e.~$\le0.05$\,pc). 

Both spectral index and diameter trends support the hypothesis that variability is greater in younger \uchii\ regions. Positive spectrum \uchii\ regions are among the youngest as they evolve from optically thick to optically thin emission \citep[e.g., \hchii\ regions][]{Yang2019MNRAS4822681Y,Yang2021AA645A110Y}. Similarly, even though young \hii\ regions may ``collapse'' and change brightness as they evolve \citep{Peters2010a}, the overall picture is one of expansion over time. Our variable \uchii\ region sample and the Top7 variable \uchii\ regions are amongst the youngest \uchii\ regions to be discovered. This is consistent with the theoretical picture presented by \citet{Peters2010a,Peters2010b}, where \uchii\ regions undergo frequent changes in brightness early on in their evolution as they fluctuate between gravitationally trapped and extended states. 
Alternatively, as young \hii\ regions evolve, their turnover frequency between optically-thick and optically-thin regimes decreases \citep{Kurtz2005IAUS,Yang2021AA645A110Y,Patel2023MNRAS5244384P}. Consequently, it is possible that the flux density at an optically thick frequency increases, resulting from an evolution transition to optically thinner conditions. 
However, the flux variations due to this transition of young \hii\ regions from the classical model are tiny at such a short timescale of $\sim$8\,yr.

  \begin{figure*}[!htp]
 \centering
\begin{tabular}{cc}
\includegraphics[width = 0.45\textwidth]{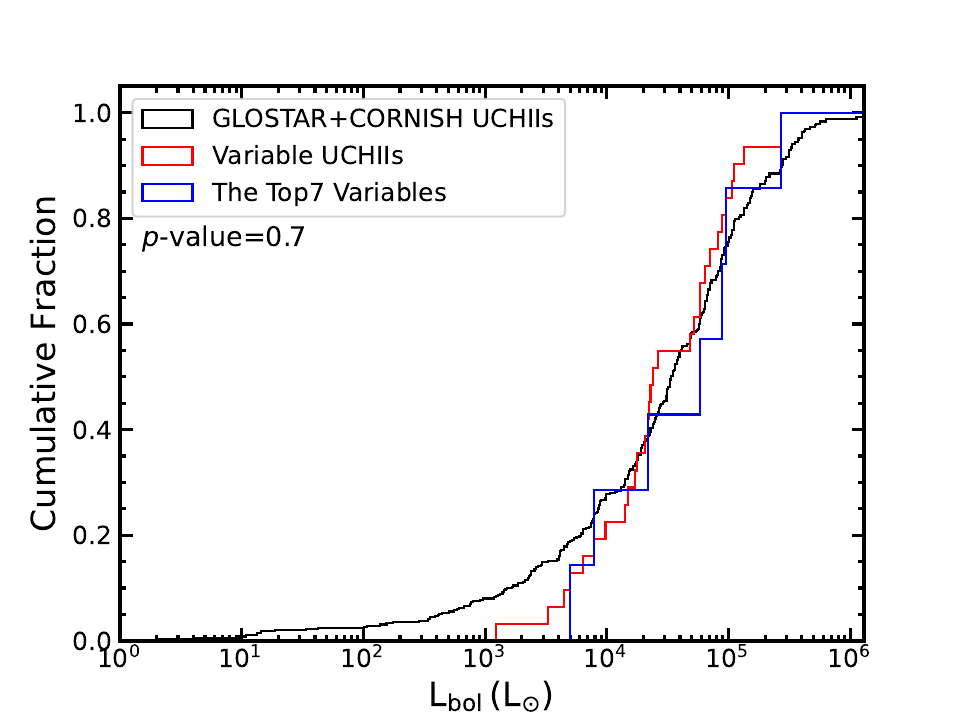} &
\includegraphics[width = 0.45\textwidth]{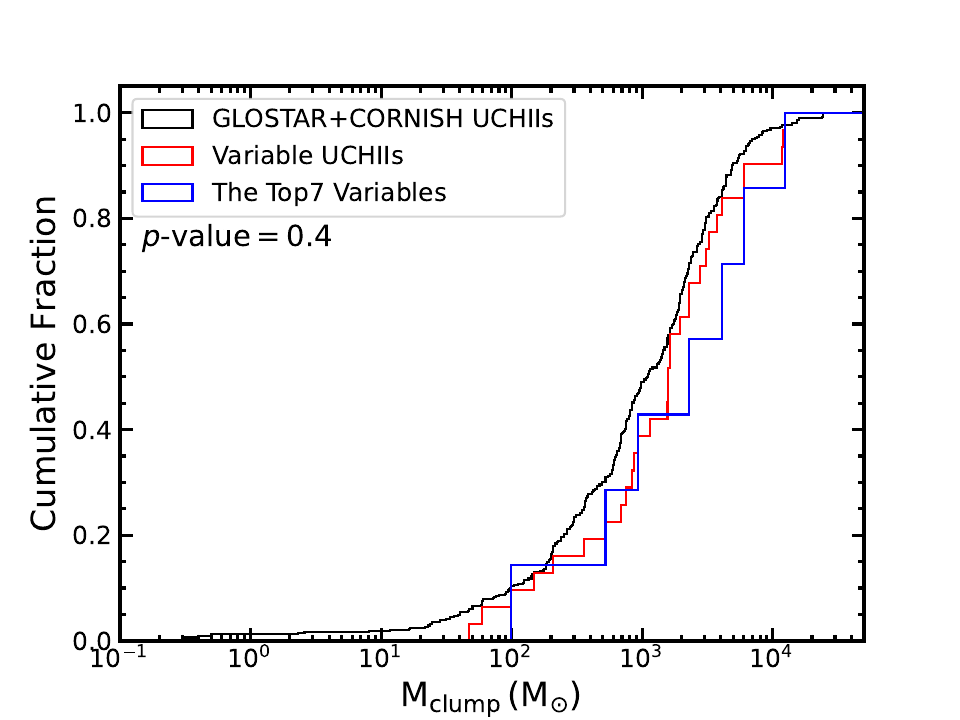} \\
\includegraphics[width = 0.45\textwidth]{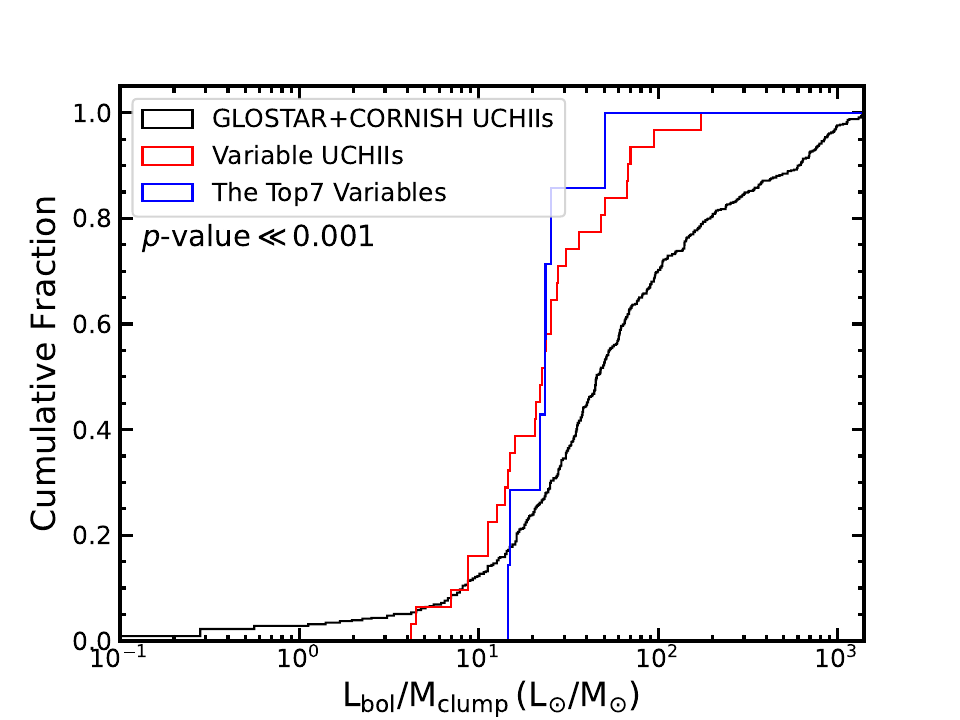} &
\includegraphics[width = 0.45\textwidth]{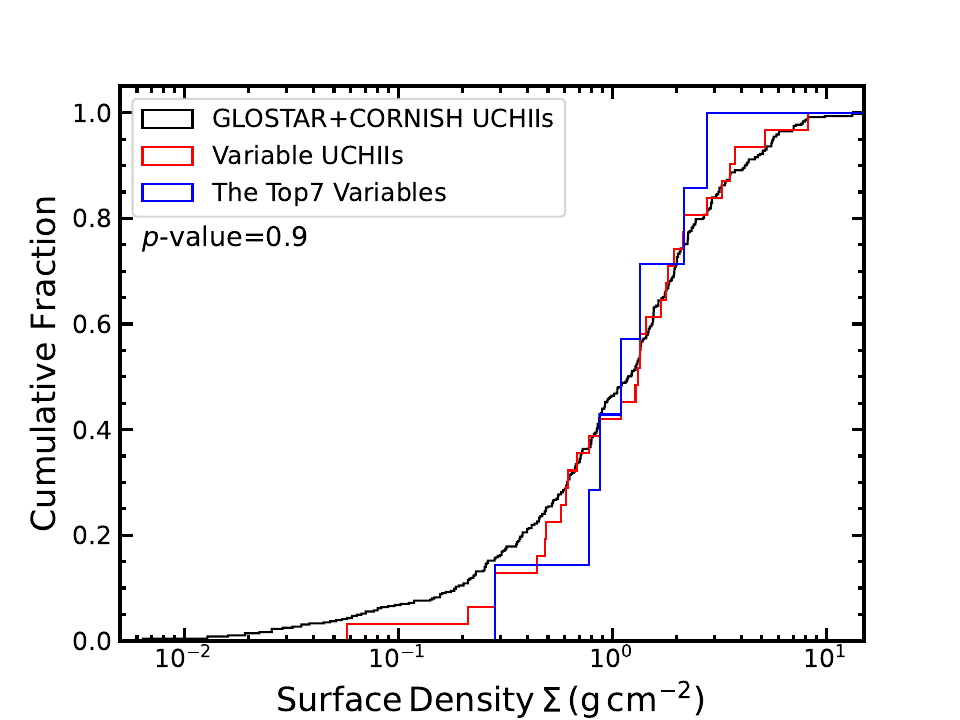} \\
\includegraphics[width = 0.45\textwidth]{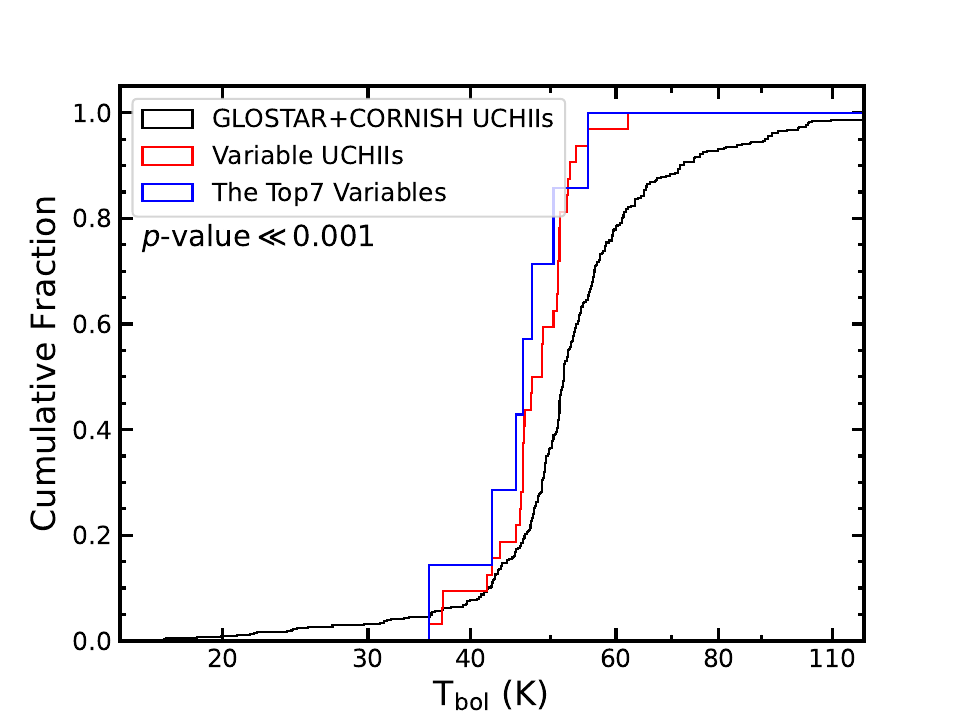} &
\includegraphics[width = 0.45\textwidth]{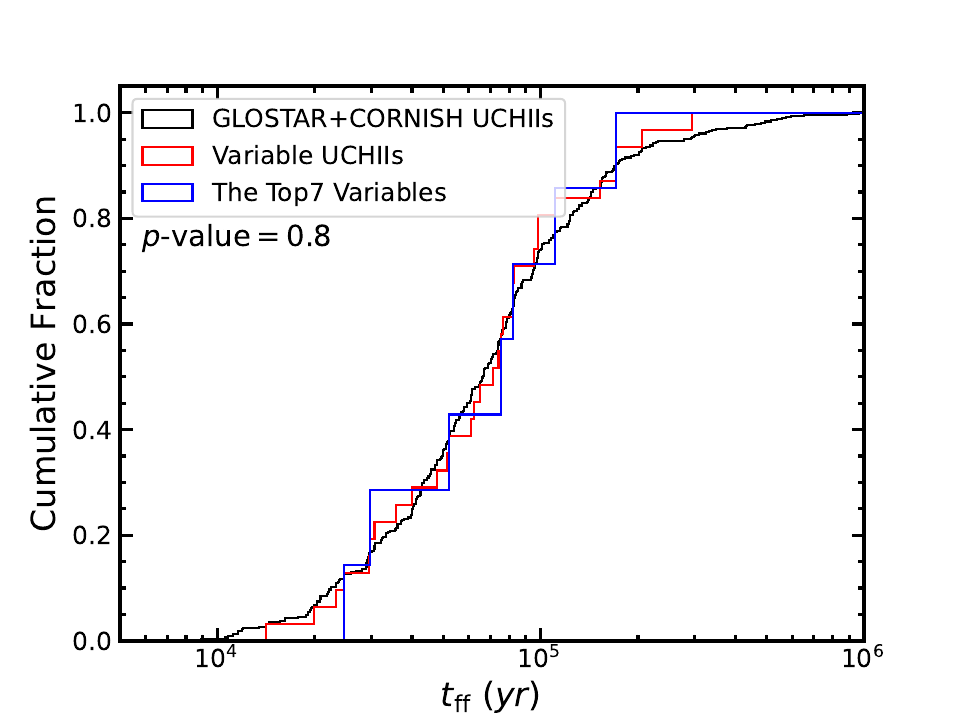} \\
\end{tabular}
\caption{Cumulative distributions of the clump properties such as bolometric luminosity $L_{\rm bol}$, mass $M_{\rm clump}$, luminosity-to-mass ratio $L_{\rm bol}/M_{\rm clump}$, surface density $\Sigma$, bolometric temperature $T_{\rm bol}$, and the clump free-fall times $t_{\rm ff}$ for the natal clumps of the Top7 variables (blue) and variable \uchii\ regions (red) of this work, as well as the GLOSTAR \uchii\ regions (black) and CORNISH \uchii\ regions. }  
 \label{fig:distr_var_uchii_vs_uchii}
 \end{figure*}

  \begin{figure}[!htp]
 \centering
\includegraphics[width = 0.45\textwidth]
{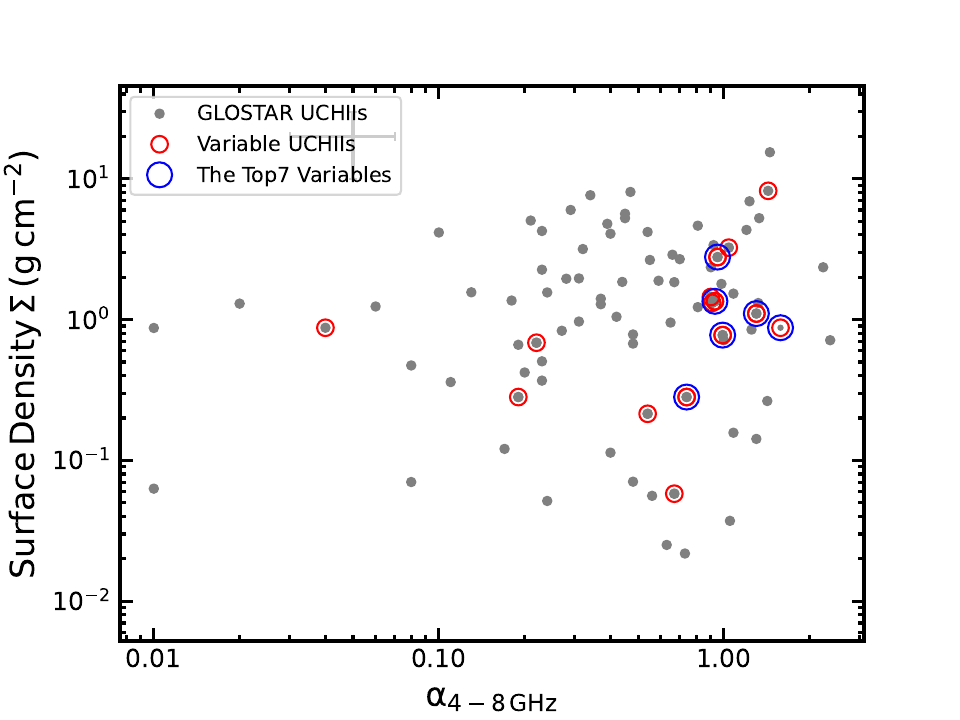} \\
\includegraphics[width = 0.45\textwidth]
{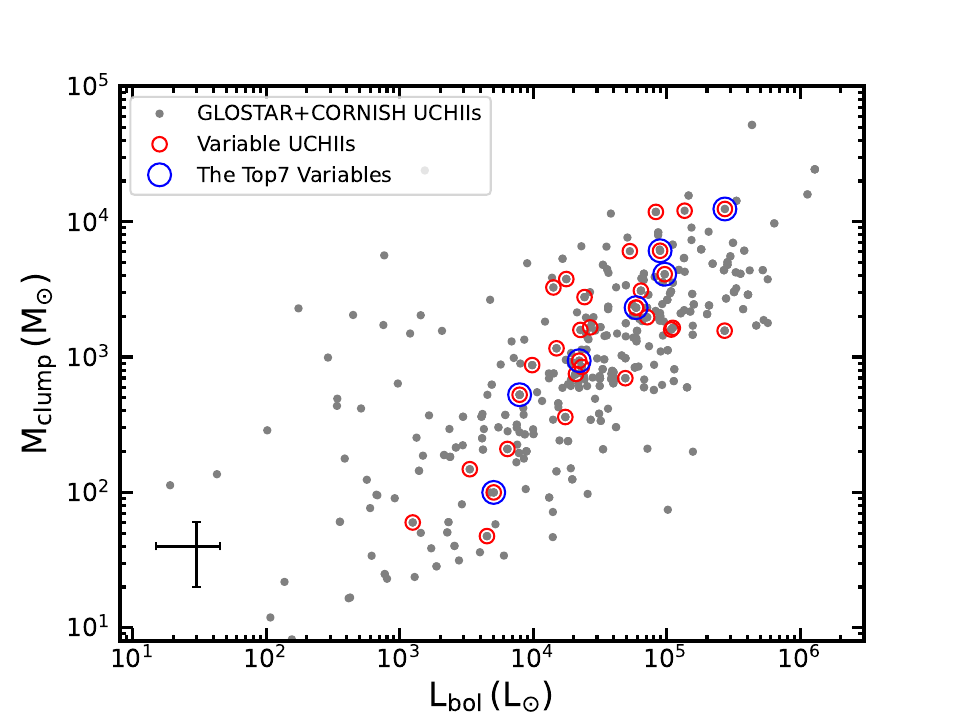} \\
\includegraphics[width = 0.45\textwidth]{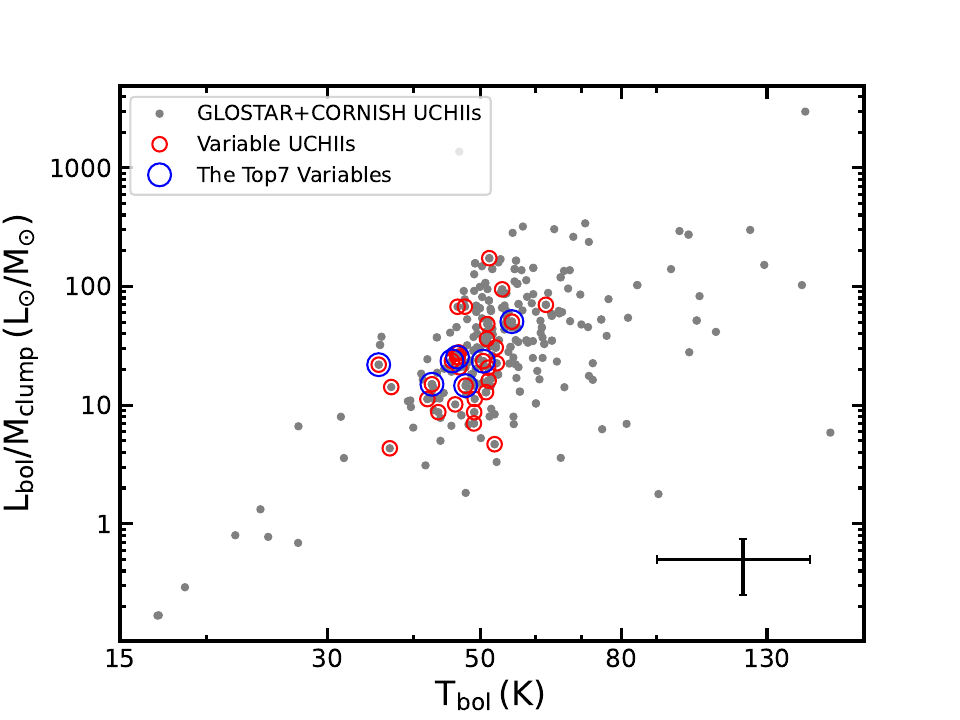} \\
\caption{Top panel: The spectral index $\alpha_{4-8\,GHz}$ \uchii\ regions vs. the surface densities of their natal clumps for the variables (red circles) sample, and the Top7 (blue circles) (with 6 having reliable $\alpha$ in Table\,\ref{tab:7_most_var_HII}), and the total \uchii\ regions (grey dots) from GLOSTAR and CORNISH. 
Middle panel: The bolometric luminosity vs. the clump mass of the natal clumps for the variable \uchii\ regions, the Top7, and the total \uchii\ regions of CORNISH and CORNISH. 
Bottom panel: The bolometric temperature vs. the luminosity-to-mass ratio
of the natal clumps for the variable \uchii\ regions, the Top7 variable \uchii\ regions, and the GLOSTAR and CORNISH \uchii\ regions. 
The error bars in each figure correspond to a typical uncertainty of these parameters, such as 0.2 for $\alpha$ \citep{Yang2023AA680A92Y}, 25\% for $T_{\rm bol}$, 50\% for $L_{\rm bol}$,  $M_{clump}$, and $L_{\rm bol}/M_{\rm clump}$ \citep{Urquhart2018MNRAS4731059U,Urquhart2022MNRAS3389U}.}  
 \label{fig:clumps_properties_of_variables}
 \end{figure}

\subsection{Clump properties}
\label{sect:clump_prop}

In Figs.\,\ref{fig:distr_var_uchii_vs_uchii} and \ref{fig:clumps_properties_of_variables}, we present the distributions of the clump properties such as the bolometric luminosity $ L_{\rm bol}$, clump mass $M_{\rm clump}$, luminosity-to-mass ratio $L_{\rm bol}/M_{\rm clump}$, bolometric temperature $T_{\rm bol}$, and surface densities $\sum$, and the clump free-fall times $t_{\rm ff}$ \citep{Urquhart2018MNRAS4731059U,Urquhart2022MNRAS3389U} of the natal clouds for the variable \uchii\ regions (red), the Top7 variable \uchii\ regions (blue), and the total \uchii\ region sample (back/grey) from CORNISH and GLOSTAR.
Fig.\,\ref{fig:clumps_properties_of_variables} suggests that the natal clumps of variable \uchii\ regions do not exhibit any distinctive properties. 
This is also supported by the cumulative distributions of the clump properties such as $L_{\rm bol}$, $M_{\rm clump}$,  $T_{\rm bol}$, $\sum$, and $t_{\rm ff}$ in Fig.\,\ref{fig:distr_var_uchii_vs_uchii}, with $p$-values $\geq$ 0.4 from the K-S tests between the two samples of variable \uchii\ regions and the total \uchii\ region sample. 
However, significant differences are found in the $L_{\rm bol}/M_{\rm clump}$ and $T_{\rm bol}$ with $p$-value $\ll$ 0.001 between the two samples.
The natal clumps of the variable \uchii\ regions are found to have lower values of $L_{\rm bol}/M_{\rm clump}$ and $T_{\rm bol}$, indicating that they are in earlier evolutionary stages with a lower bolometric temperature, as the evolutionary stages of clumps discussed in \citet{Konig2017AA599A139K} and \citet{Urquhart2022MNRAS3389U}, compared to the natal clumps of the whole sample of \uchii\ regions. 
This suggests that the variability observed in \uchii\ regions is not related to the physical properties of clumps, but is significantly linked to the evolutionary stages of clumps, preferably occurring in the early stages of massive star formation.

Among the 38 variable \uchii\ regions, outflow searches have been conducted in 26 of their natal clumps in \citet{Yang2018ApJS2353Y, Yang2022AA658A160Y}, 15 (58\%=15/26) of them are found to be associated with molecular outflows. 45\% (17/38) of them are found to be associated with methanol masers from the GLOSTAR survey \citep{Nguyen2022AA666A59N}. 
Among the Top7 variable \uchii\ regions, 5 of their natal clumps with outflow searching, 60\% (3/5) are found to be associated with molecular outflow clumps in \citet{Yang2018ApJS2353Y, Yang2022AA658A160Y}, and 88\% (6/7) are found to be associated with methanol maser emissions from the GLOSTAR survey \citep{Nguyen2022AA666A59N}. 
From the total GLOSTAR \hii\ regions sample, we found a smaller percentage of $\sim 37\% $\,(146/390) associated with methanol masers and a similar fraction of $\sim 56\%$\,(151/272, where 272 have their natal clumps with outflow searching) associated with outflows, compared to the variable \uchii\ region sample in this study. 
Given that outflows and methanol masers are indicative of early evolutionary stages of massive star formation \citep[e.g.,][]{Menten1991ApJ380L75M,Chen2011ApJS1969C,Qiu2009ApJ69666Q,Yang2022AA658A192Y,Luo2023ApJ952L2L,Mai2024ApJ961L35M}, this provides further evidence that the variable \uchii\ regions sample and the Top7 are in the very early stage of \uchii\ regions. This is also supported by the fact that all of variable \uchii\ regions in this study are found to meet the size criterion of \hchii\ regions, and some of which have been identified as \hchii\ regions in \citet{Yang2021AA645A110Y}.

Additionally, all of the natal clumps exhibit no infrared variability, with one exception of the natal clump, AGAL036.878-00.474, which displays infrared variability as observed by \citet{Lu2024ApJS27244L} based on the NEOWISE database. A connection between the radio flux density and the natal clumps would be interesting for future exploration. 
  
\subsection{The most variable \uchii\ region}

From Fig.\,\ref{fig:alpha_vs_fluxratio_hii}, Table\,\ref{tab:7_most_var_HII} and Fig.\,\ref{fig:38_variables_radio_images}, we observe that the \uchii\ region G010.4629$+$00.0299 exhibits the highest variability, with its flux increasing from 2.44\,mJy in CORNISH to 11.15\,mJy in GLOSTAR, giving a more than fourfold increase over a timescale of 8 years. 
Flux changes have been reported on several \hchii\ regions and \uchii\ regions over the timescales of several years \citep[][]{Franco2004ApJ604L105F,vanderTak2005AA431993V,Galvan_Madrid2008ApJ674L33G,DePree2015ApJ815123D,Rivilla2015ApJ808146R,Brogan2018ApJ86687B,Hunter2018ApJ854170H,DePree2018ApJ863L9D,DePree2020AJ160234D}, showing typical variations of approximately 20-50\%, with no reported examples of flux variations exceeding 100\%.   
This makes it the highest variability reported for \hii\ regions so far, over a short timescale of 8\,years. 

As discussed in Sect.\,\ref{sect:clump_prop}, no extreme physical properties have been observed in the natal clump of this most variable source. 
However, its natal clump displays high maximum outflow velocities, with velocities twice as high as the mean value of the sample of 1192 outflow clumps, as presented in Figure 5 of \citet{Yang2022AA658A160Y}. 
The most variable G010.4629$+$00.0299, together with the second most variable  G010.4724$+$00.0274 classified as an \hchii\ region in \citet{Yang2021AA645A110Y}, are located in the cluster environment of the complex star formation region W31, with detected various molecules  
\citep{Hatchell1998AAS13329H,Ren2014AA567A40R}, $\rm H_{2}O$ maser \citep{Hofner1996AAS120283H}, OH masers \citep{Braz1987AA18119B}, and methanol maser \citep{Green2010MNRAS409913G,Yang2019ApJS24118Y,Nguyen2022AA666A59N}. 
The cluster and clumpy environment observed at submillimeter ALMA \citep{Gieser2023AA674A160G}, mid-infrared \citep{Pascucci2004AA426523P}, and radio \citep{Yang2021AA645A110Y} with angular resolutions of $\lesssim 1.0\arcsec$, and the clump-scale high-velocity outflows detected by CO \citep{Lopez_Sepulcre2009AA499811L,Yang2022AA658A160Y}, support that the variations of \uchii\ regions preferably occur in young and active massive star-forming regions, which might be connected to the unstable and clumpy accretion flows.

\section{Conclusion and Summary}
\label{sect:conclusions}

By comparing the CORNISH and GLOSTAR 5GHz surveys, we have identified a sample of 38 variable \uchii\ regions, the largest sample of variable \uchii\ regions identified to date. The high angular resolution of CORNISH \citep{Hoare2012PASP} combined with the broadband sensitivity of GLOSTAR \citep{Brunthaler2021AA651A85B} and the detailed analysis of natal molecular clumps \citep{Urquhart2018MNRAS4731059U} has allowed us to examine the relationship between variability and physical properties for the first time. Our main conclusions are as follows:

\begin{enumerate}

     \item  Within the whole sample of 38 variable \uchii\ regions we see that all sources with a decrease in their 5 GHz flux density have a negative radio spectral index. We speculate that these sources may be contaminated by non-thermal radio jets. \uchii\ regions showing flux increases have a mix of positive and negative spectral indices.\\
     
     \item There is a significant correlation between the spectral index and the CORNISH-GLOSTAR flux ratio. \uchii\ regions with steeper spectral indices (both positive and negative) show greater variability in their radio flux density. Variable \uchii\ regions also have significantly smaller diameters than the non-variable sample. When considering distributions of spectral index and diameter, we are confidently able to reject the null hypothesis that variable \uchii\ regions are drawn from the same parent sample as non-variable \uchii\ regions. We hypothesize that the most variable sources are also the youngest \hii\ regions, with diameters and spectral indices consistent with hypercompact \hii\ regions. This is consistent with the \citet{Peters2010a,Peters2010b} models where the largest changes in brightness occur at the earliest stages.\\
     
     \item When considering the most variable \uchii\ regions (the Top7 subsample), we also find that they have positive spectral indices, increase in flux density, and small diameters. This may be due to the spectral evolution of the sample as the \hii\ regions transition from optically thick to optically thin at 5 GHz. This evolution would result in a gradual brightening of the 5 GHz flux density.\\

     \item There is no significant correlation between the physical properties of the natal molecular clumps hosting the \uchii\ regions and their variability. However, we do find a significant difference in evolutionary indicators of the clump ($T_{\rm bol}$ and $L_{\rm bol}$/$M_{\rm clump}$) whereby the younger clumps host the most variable \hii\ regions. This is consistent with our finding that younger \hii\ regions are the most variable.\\

     \item The overall occurrence of variability within our \uchii\ region sample ($\sim$44\%) is larger than that predicted $\sim 10\%$ by the models in \citet{Galvan_Madrid2011MNRAS4161033G}, suggesting that flux variation in \uchii\ regions is significantly more common than predictions. 

\end{enumerate}

With the available observational data, we are able to identify a large sample of variable \uchii\ regions on decade timescales and uncover significant relations between variability and the evolutionary stage of the \uchii\ regions and their natal molecular clumps. However, it is clear that a broadband monitoring campaign of \uchii\ regions is required to understand the mechanism behind their variability, uncover details of their spectral evolution, and hence constrain accretion models of the massive star formation process.

\section{Data Availability}
\begin{figure*}[!h]
\centering
\begin{tabular}{cc}
\includegraphics[width=0.45\textwidth, trim= 0 0 0 0,clip]{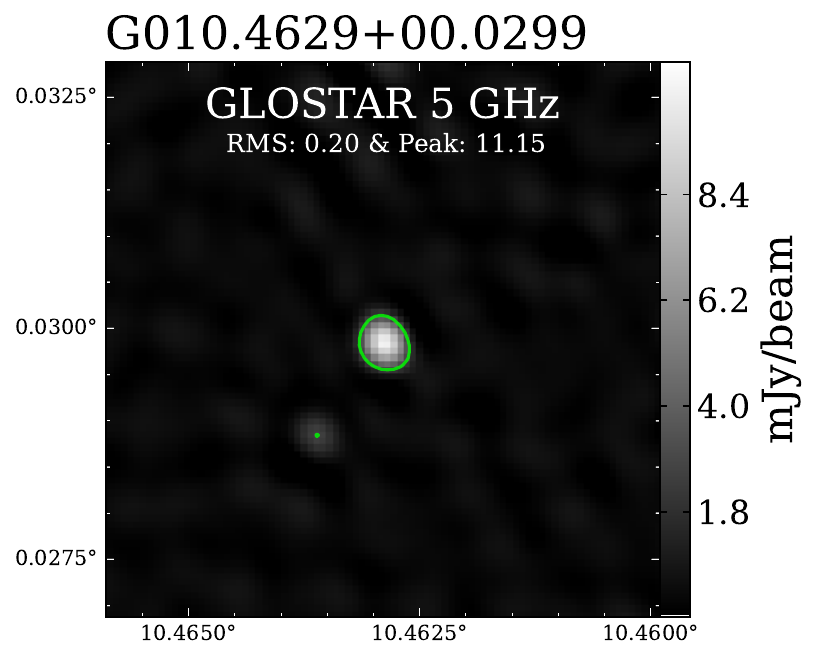}  &  \includegraphics[width=0.45\textwidth, trim= 0 0 0 0,clip]{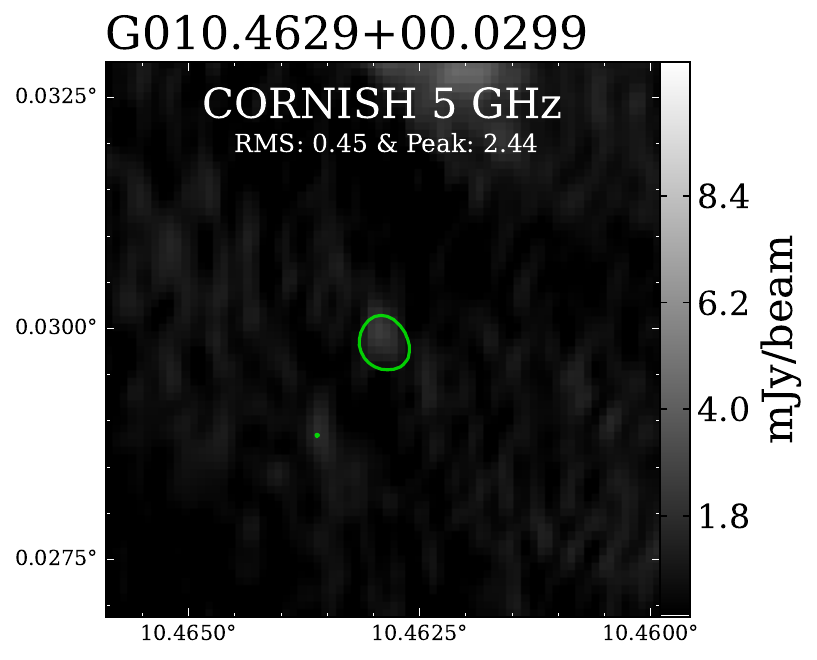} \\
 \includegraphics[width=0.45\textwidth, trim= 0 0 0 0,clip]{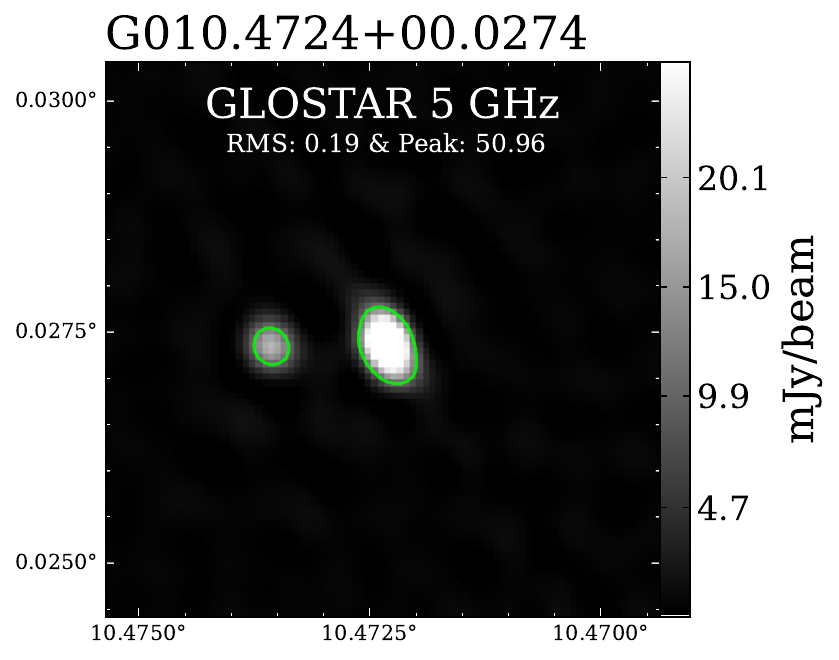}  &  \includegraphics[width=0.45\textwidth, trim= 0 0 0 0,clip]{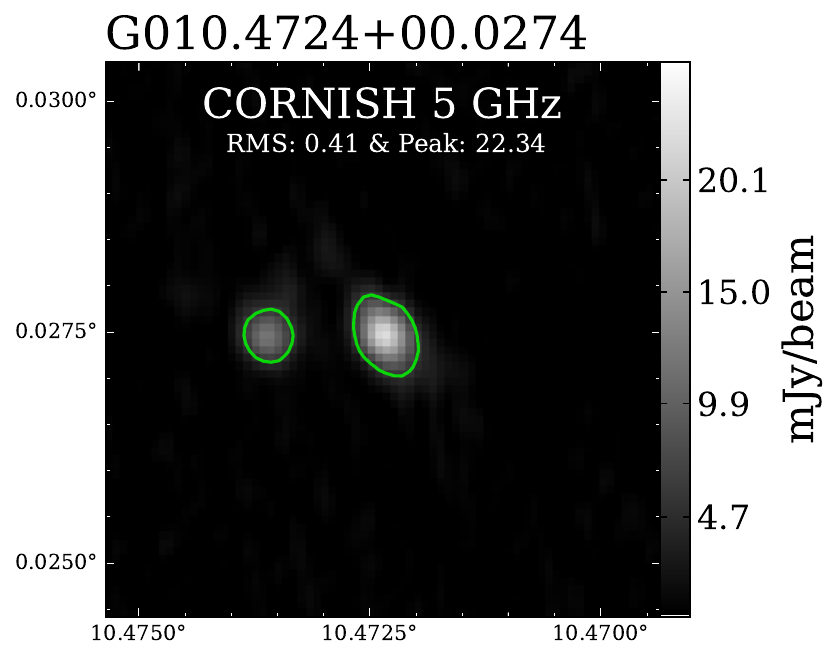} \\
 \includegraphics[width=0.45\textwidth, trim= 0 0 0 0,clip]{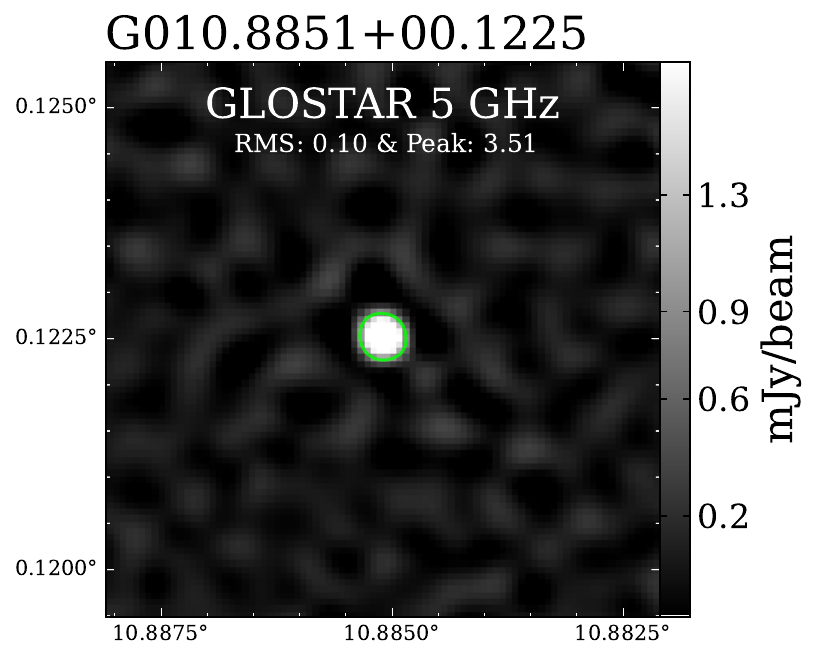}  &  \includegraphics[width=0.45\textwidth, trim= 0 0 0 0,clip]{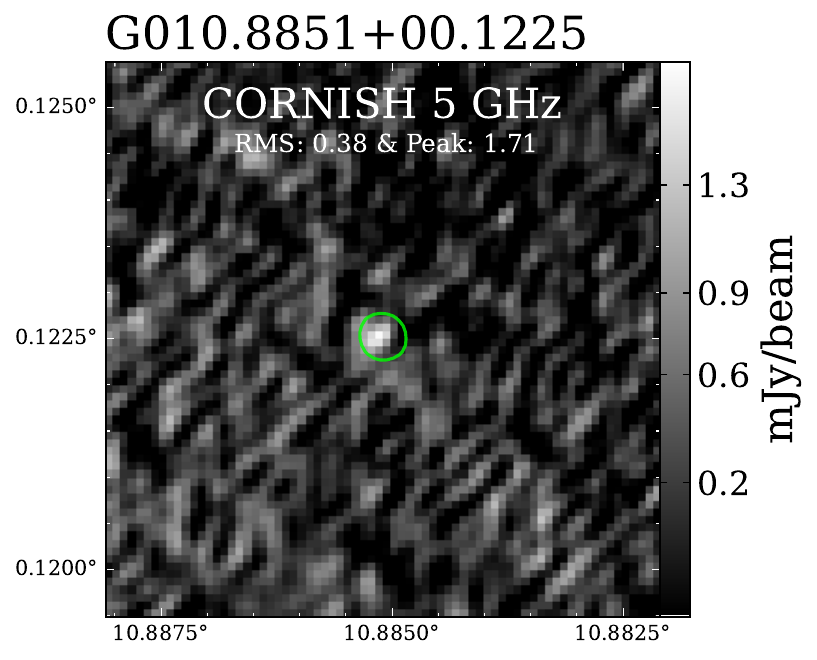} \\
\end{tabular}
\caption{ The GLOSTAR and CORNISH 5 GHz images for the 38 variable sources identified in this work. The beam size is 1.5\arcsec for each image in the two surveys. Each figure is  centered at the position of the identified \hii\ region. Only a small portion of the sample is presented here, and the radio images for the full sample of 38 variable \uchii\ regions are available in electronic form at the
Zenodo via \url{https://zenodo.org/records/13906200}. }
\label{fig:38_variables_radio_images}
\end{figure*}

The data underlying this article are available in the article and in its online supplementary material. The full version of Table\,\ref{tab:7_most_var_HII} is
available on CDS. The full version of Figure\,\ref{fig:38_variables_radio_images} is available on in electronic form at the Zenodo via \url{https://zenodo.org/records/13906200}.

\begin{acknowledgements}
We thank the anonymous referee for providing useful suggestions and comments, which helped to improve the manuscript. 
AYY acknowledges the support from the National Key R$\&$D Program of China No. 2023YFC2206403 and National Natural Science Foundation of China (NSFC) grants No. 12303031 and No. 11988101. 
DL is a New Cornerstone investigator. 
YG is supported by the Strategic Priority Research Program of the Chinese Academy of Sciences, Grant No. XDB0800301. 
TCW and DL acknowledge support from the International Partnership Program of Chinese Academy of Sciences, Program No.114A11KYSB20210010. 
The National Radio Astronomy Observatory is a facility of the National Science Foundation, operated under a cooperative agreement by
Associated Universities, Inc. 
This research made use of Astropy (http://www.astropy.org), a community developed core Python package for Astronomy \citep{Astropy2013AA558A33A,Astropy2018AJ156123A}.
\end{acknowledgements}

\bibliographystyle{aa}
\bibliography{ref}

%
%

\end{document}